%% file: DecoupledTCP.tex
\documentclass[conference, twocolumn, 10pt]{IEEEtran}
\ifCLASSINFOpdf
  \usepackage[pdftex]{graphicx}
   \graphicspath{{..1/pdf/}{../jpeg/}}
   \DeclareGraphicsExtensions{.pdf,.jpeg,.png}
\else
   \usepackage[dvips]{graphicx}
   \graphicspath{{../eps/}}
   \DeclareGraphicsExtensions{.eps}
\fi

\usepackage{cite}
\usepackage{comment}
\usepackage[cmex10]{amsmath}
\usepackage{amsfonts, amsmath, amsthm, amstext, mathrsfs, graphicx, color, url}
\usepackage{array}
\usepackage{mdwmath}
\usepackage{mdwtab}
\usepackage{algpseudocode}
\usepackage{algorithm}
\usepackage{url}
\usepackage{setspace}
\usepackage{footnote}
\usepackage[skip=0pt]{caption}
\usepackage{float}
\usepackage{epstopdf}
\usepackage{epsfig, amsmath,amssymb,epsf,cite,subfigure,scalefnt,multirow,array,setspace}
\newcommand\subparagraph{%
  \@startsection{subparagraph}{5}
  {\parindent}
  {3.25ex \@plus 1ex \@minus .2ex}
  {-1em}
  {\normalfont\normalsize\bfseries}}
\makeatother
\usepackage{titlesec}
\let\subparagraph\relax
\titlespacing*{\section}
{0pt}{2pt}{2pt}
\titlespacing*{\subsection}
{0pt}{2pt}{2pt}
\titlespacing*{\subsubsection}
{0pt}{2pt}{2pt}

\input{Macros.tex}

\begin{document}

\title{TCP Decoupling for Next Generation Communication System }

\author{
\IEEEauthorblockN{
Xiaohui Chen\IEEEauthorrefmark{1},
Xiaowei Qin\IEEEauthorrefmark{1},
Li Chen\IEEEauthorrefmark{1},
Huarui Yin\IEEEauthorrefmark{1},
Weidong Wang\IEEEauthorrefmark{1},
Guo Wei\IEEEauthorrefmark{1},\\
\{cxh,qinxw,chenli87,yhr,wdwang,wei\}@ustc.edu.cn\\
Tianyi Zhang\IEEEauthorrefmark{1}\IEEEauthorrefmark{2},
Yanbin Liu\IEEEauthorrefmark{1},
Ting Zhu\IEEEauthorrefmark{1},
Hailun Liu\IEEEauthorrefmark{1},\\
tianyi.zhang@ufl.edu \{viper, zhuting, lhl827\}@mail.ustc.edu.cn
\\
}

\IEEEauthorblockA{\IEEEauthorrefmark{1}Department of Electronic Engineering and Information Science, University of Science and Technology of China, hefei, China}
\IEEEauthorblockA{\IEEEauthorrefmark{2}Department of Electrical and Computer Engineering, University of Florida, Florida, USA}
}
\maketitle

\begin{abstract}
In traditional networks, interfaces of network nodes are duplex. But, emerging communication technologies such as visible light communication, millimeter-wave communications, can only provide a unidirectional interface when cost is limited. It's urgent to find effective solutions to utilize such new unidirectional communication skills.
Decoupling implies separating one single resource to two independent resources. This idea can be applied at physical layer, link layer, network layer, even transport layer. TCP decoupling is an end to end solution provided at transport layer. With decoupled TCP, two distinct unidirectional path can be created to meet the requirements of reliable information transfer. However, it is not an easy task to decouple a bidirectional logical path at transport layer.
In this paper, we dwell on the idea of TCP decoupling. Advantages of decoupling at transport layer are analyzed also. In addition, an experiment is carried out to figure out how to implement a decouple TCP. Our results show decoupling at transport layer is possible and the modified protocol is available.
\end{abstract}

\IEEEpeerreviewmaketitle

\section{Introduction}
\vspace{.6em}
The Transmission Control Protocol (TCP) is one of the main transfer layer protocols. In order to provide reliable delivery of a data stream, a bi-directional path has been assumed. In other words, the two directions have been coupled in a network interface. So it's nature to design TCP on a bidirectional logical channel. A terminal can send data packets from the logical channel. Meanwhile, acknowledgement of data packets can be received through the same channel. All in all, the traditional TCP and its variants run over a bi-directional network interface.

Nowadays, terminals equipped with multiple network interfaces are becoming commonplace. There are numerous solutions to support concurrent data transfer over multiple interfaces~\cite{mp-2016}. For example, aggregation skills are implemented at link layer~\cite{dual-2016}~\cite{dual-2017}. Tunneling mechanisms are designed in IP level~\cite{mp-2016}. In contrast, Multipath TCP can balance a single TCP connection across multiple interfaces at transport layer~\cite{mptcp-2013}. Although various ideas introduce different methods for multiple path transfer, a path is assumed as a bi-directional channel.

However, not all emerging communication technologies hold the bidirectional feature. In some circumstance, it is easy to implement a unidirectional channel while bi-directional channel is difficult. For example, visible light communications~\cite{vlc-2016}, massive MIMO millimeter-wave communications, high capacity link from access point to terminal is more practical than the opposite direction. Let's imagine the possible scenarios in future 5G communication system. There would be traditional 3G/4G bi-directional channel, high rate millimeter-wave downlink channel, short range WiFi bi-directional channel and indoor VLC downlink channel. How to take advantages of all these unidirectional or bidirectional channels? It's sure that the traditional TCP cannot meet such demand unless link layer aggregating integrate all these interfaces together. In other words, a bidirectional path has to be established. Are there some ways to utilize multiple unidirectional channel at transfer layer? Certainly, TCP decoupling make it possible. The idea behind TCP decoupling lies in that reliable transfer can be realized not only on a bidirectional path, but also on two unidirectional paths with opposite transfer directions.

Decoupling is not a new idea in communication systems. Some related researches can be found in the past literatures. For example, downlink and uplink decoupling at physical layer in cellular network~\cite{dedu-2016}, Decoupling downlink to high frequency band while uplink is limited to low frequency is also under consideration~\cite{dedu-2016}~\cite{5g-2016}, decoupling of control and data at link layer~\cite{aggr-2013}. TCP Decoupling is a solution at transport layer. It's well known that TCP guarantees reliable information transfer through packets switch at two directions. After data packets are transmitted from one node to another, acknowledgements of these packets are transmitted on opposite direction. Decoupling at transport layer means that the path of data packets transfer and the path of acknowledgements transfer can be decoupled.

In other words, TCP decoupling means there are two different unidirectional logical paths between source node and destination node. The end to end path decoupling introduces more flexibility to utilize the communication resources in network, helps overcoming the weakness of asymmetric links, facilitates to aggregate heterogeneous links in network, especially for wireless network. These advantages make it possible to defeat traditional TCP in some scenarios. Potential application scenarios include high rate visible light communication, millimeter-wave communications, deep space communications, etc.

The remainder of this paper is organized as follows. In section II, we emphasize on the difference and relation between link and path, which is important to understand TCP decoupling. Advantages of decoupled TCP are analyzed in section III. The procedure of TCP path decoupling is provided in section IV. In section V, we propose a design of decoupled TCP, experiment and its results are shown also. Open issues and future works are discussed in section VI.

\vspace{.6em}
\section{From link decoupling to path decoupling}
\begin{figure}[!t]
 \centering
 \includegraphics[scale = 0.3]{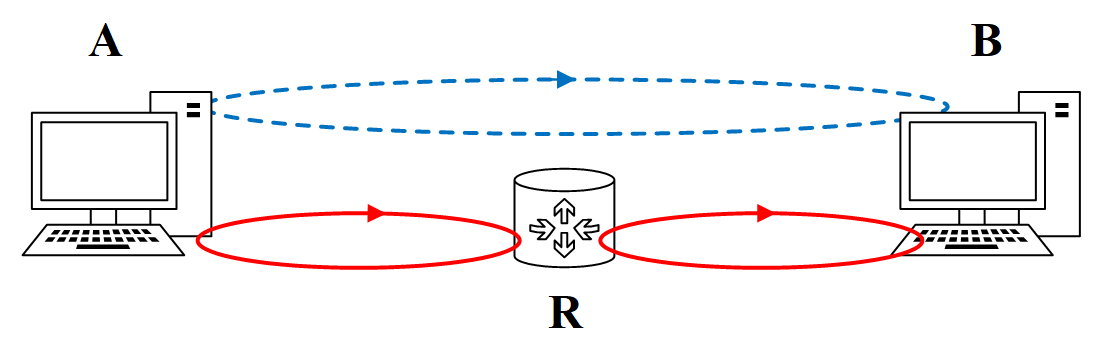}
\caption{Comparison of link and path}
 \label{fig1}
\end{figure}

\vspace{.6em}
  %\vspace{-.5em}
In telecommunication, a link refers to a set of electronics assemblies, consisting of a transmitter and a receiver (two pieces of data terminal equipment) and the interconnecting data telecommunication circuit. At the same time, A path refers to a set of links, consisting of tandem links. A path can be a single link in the simplest scenario. Commonly, a path includes multiple links which are connected through routers. It is shown in Fig. \ref{fig1}, message flow (A-B) is in the presence of a router (R), while, blue dash flow is effective communication paths, red solid paths are across the actual network links.

There are at least three types of links. The first is simplex communications, most commonly meaning all communications in one direction only. The second is half-duplex communications, meaning communications in both directions, but not both ways simultaneously. The third is duplex communications, communications in both directions simultaneously. But, a path in TCP (Transmission Control Protocol) is usually assumed bidirectional. That is, a path provides data packets transfer in both directions simultaneously.

\subsection{Decoupling of link and path}
  %\vspace{-.5em}
Ever since the inception of mobile telephony, the downlink and uplink of cellular networks have been coupled, that is, mobile terminals have been constrained to associate with the same base station in both the downlink and uplink directions~\cite{dedu-2016}. New trends in network technologies, emerging novel communication technologies and mobile data usage increase the drawbacks of this constraint. The authors of ~\cite{dedu-2016} suggest to enable link decoupling in 5G (fifth generation) cellular standards. They demonstrate that decoupling can lead to significant gains in network throughput, outage, and power consumption at a much lower cost compared to other solutions that provide comparable or lower gains.

\begin{figure*}[!t]
 \centering
 \includegraphics[scale = 0.4]{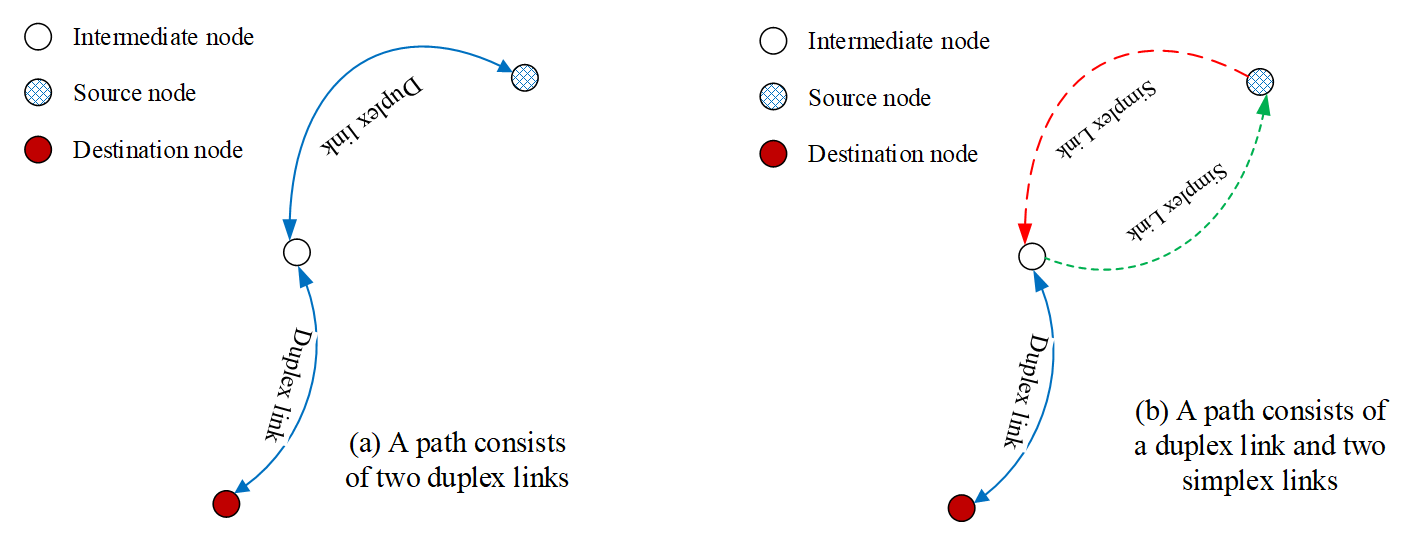}
\caption{A path consists of duplex or simplex links}
 \label{fig2}
\end{figure*}

Considering a link is an element in a path, link decoupling introduces the possibility of path decoupling. A path consists of duplex links is shown in Fig. \ref{fig2} (a), whose two directions are coupled. A path consists of a duplex link and two simplex links is shown in Fig. \ref{fig2}(b), which is called decoupled because the two directions of data transfer are mapped to different paths. A decoupled TCP refers to TCP transfer over decoupled path, similar to scenario in Fig.\ref{fig2}(b).

\subsection{What happened after path decoupling}
\begin{figure*}[!t]
 \centering
 \includegraphics[scale = 0.5]{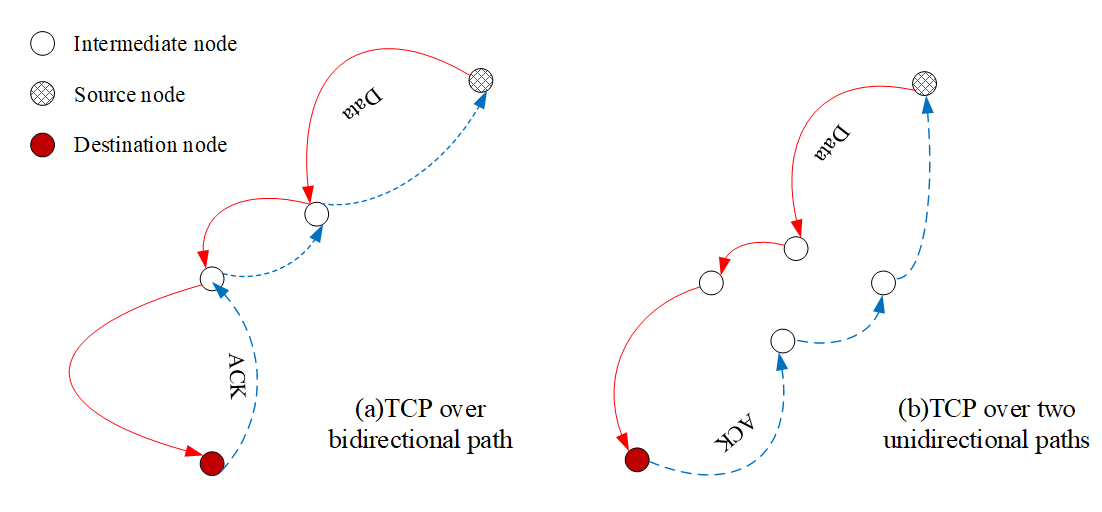}
\caption{Comparison of coupled TCP and decoupled TCP}
 \label{fig3}
\end{figure*}

  %\vspace{-.5em}
Information transfer from one node to another implies a certain direction. We define this direction as the forward direction (of information transfer), and, the opposite direction is defined as reverse direction. Commonly, message which carries information is sent on the forward direction. If acknowledgement is needed, ACK of this message would be sent on the reverse direction. Under the circumstance of TCP, data packets are sent by source node on the forward direction, at the same time, ACK packets are sent by destination node on the reverse direction.
A terminal with single bidirectional network interface distinguishes the two direction by send or receive modules. It is noteworthy that the interface used by TCP provides sending and receiving ability at the same time. The procedure of data packets sending and ACKs receiving is shown in Fig. \ref{fig3}(a). In Fig. \ref{fig3}, the solid line denotes the forward direction, and the dotted line denotes the reverse direction.
Let's assume that a terminal is equipped with two unidirectional interfaces. The one holds the ability of sending, and the other can only receiving. So the data packets and ACKs would be handled by the two interfaces, respectively, which is shown in Fig. \ref{fig3}(b). Considering that a certain interface is connected to a certain line (or wireless link), a certain interface is mapped to a certain path. Now the forward direction and the reverse direction are mapped to two unidirectional path.

\begin{figure*}[!t]
 \centering
 \includegraphics[scale = 0.3]{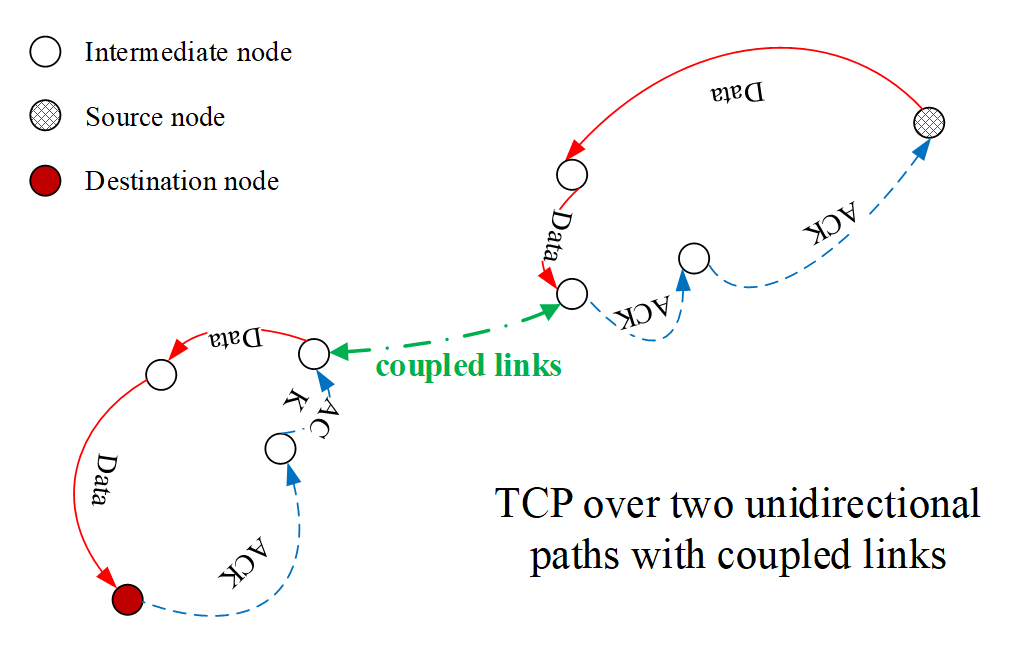}
\caption{Decoupled TCP with partial coupled links}
 \label{fig4}
\end{figure*}

Unfortunately, the topology of a real network is more complex. Each interface of terminals is configured with a default gateway. So a path in a network consists of three parts: the link between source node and the gateway of it, the link between destination node and the gateway of it, the tandem links between two gateways. In most cases, the two parts before are fixed, and, the part between two gateways is not fixed. For example, the possible path between two terminals is shown in Fig. \ref{fig4}, which consists of decoupled and coupled links. The variation is caused by the adjustment of forwarding rules of routers.

\vspace{.6em}
\section{The advantages of TCP decoupling}
\vspace{.6em}
  %\vspace{-.5em}
TCP decoupling brings new mechanism to utilize communication resources in forward or reverse directions. Link resources are allocated with coupled manner in traditional TCP. But the resource allocation with decoupled manner provides fine-granularity control. At the same time, unidirectional link can be utilized by decoupled TCP also. This benefits to aggregate heterogeneous links in next generation communication systems. Furthermore, the weakness of link asymmetries may be alleviated if extra unidirectional link is combined to reverse direction of TCP transfer.
\subsection{Decoupling leads to elaborate resource allocation}
\begin{figure*}[!t]
 \centering
 \includegraphics[scale = 0.7]{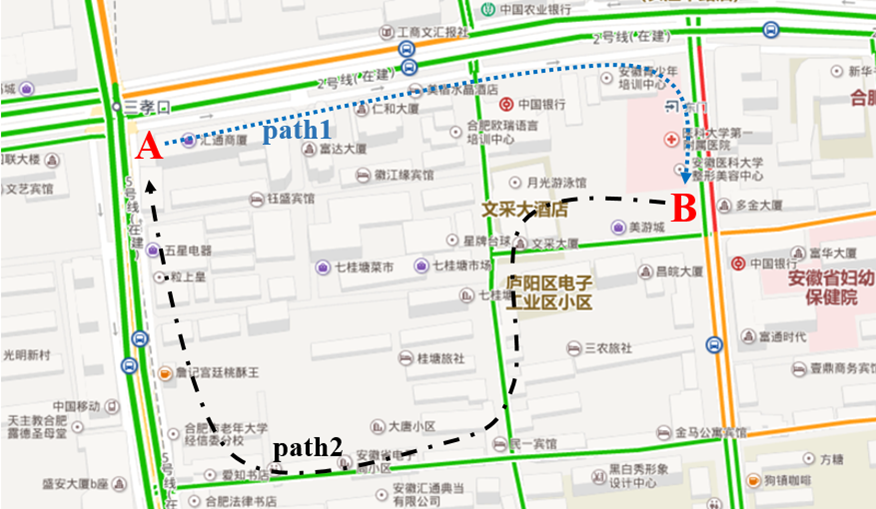}
\caption{An example of traffic navigation}
 \label{fig5}
\end{figure*}

  %\vspace{-.5em}
Let's begin with a problem of traffic navigation in Fig. \ref{fig5}. In Fig. \ref{fig5}, traffic congestion on a road is indicated by red color, and, green color means smooth traffic.If we drive from A to B, path1 (dot line in Fig. \ref{fig5}) should be chosen. But, when we drive from B to A, path2 (dash dot line in Fig. \ref{fig5}) seems more suitable. That's the advantage of decoupling!

It's similar in network topology. Assuming two terminals are equipped with two bidirectional network interfaces, then path 1 and path 2 can be established simultaneously. Obviously, path 1 is suitable for packets transport from A to B. oppositely, path 2 is suitable for packets from B to A. Recalling the transfer procedure of TCP, in which a data packet is sent on one direction and the acknowledgement of it is sent on the opposite direction. Considering the scenario of data transfer for A to B, it seems more attractive to send data packets on path 1 and to send ACK packets on path 2.
But, no matter the data packets or the ACK packets, will be sent on the same path in TCP (and all its variants). Why? It is because a bidirectional network interface is mapped to a certain bidirectional path. Although the intermediate nodes such as routers may change the transfer path of packets, it's impossible for terminals in end point. In another word, the terminals with a default gateway cannot choose different paths for data packets and ACK packets because the next hop of it can only be the default gateway.

\subsection{Decoupling facilitates to aggregate heterogeneous links}
  %\vspace{-.5em}
Aggregating of heterogeneous links is a challenging task. For example, there are various mechanisms proposed in mobile communication networks. In order to facilitate inter-working of LTE and WLAN, scenarios are discussed in 3GPP TR 22.934~\cite{tr22934}. New network elements such as ePDG (Evolved Packet Data Gateway), ANDSF (Access network discovery and selection function), TWAG/TWAP (Trusted WLAN Access Gateway/Proxy) are introduced. Furthermore, LWA (LTE-WLAN aggregation), introduced in 3GPP Release 13, is another area where protocols will be enhanced in Release 14. Release 13 supports LTE-WLAN aggregation for the DL. 3GPP Release 14 allows aggregation for the UL as well. Additional information collection and feedback (e.g., better estimation of available WLAN capacity) as well as automatic neighbor relation procedures are to be introduced to improve performance.

In fact, wireless links are half duplex in nature. The downlink and uplink directions are supported at link layer to meet the requirements of upper layer protocols. At the same time, decoupling is not a new idea in mobile communication systems. The inherent absence of duplexing for wireless links brings difficulties in deployment. Ever since the inception of mobile telephony, the downlink and uplink of cellular networks have been coupled, that is, mobile terminals have been constrained to associate with the same base station in both the downlink and uplink directions~\cite{dedu-2016}. Due to the advantages provided by link decoupling, more and more attention are paid to downlink/uplink decoupling mechanisms. For example, decoupling of control link and data link to different frequency bands is proposed in Phantom Cell~\cite{aggr-2013}. Decoupling downlink to high frequency band while uplink is limited to low frequency is also under consideration ~\cite{dedu-2016}~\cite{5g-2016}.

With the development of different access wireless technologies, heterogeneous links should be taken into account. From the view of access network, downlink and uplink stand for two categories of resource. Some links are easy to deploy in downlink direction, meanwhile, deployment in the opposite direction is very hard. Typical examples include high rate millimeter-wave for access, high rate VLC channel in indoor scenarios, and satellite channel in space communications. Aggregating such variety links at link layer is still challenging. Although the aggregating skills at network layer show enough flexibilities, these schemes need to change network elements, which introduce high cost and low compatibility. From the view of compatibility, end to end aggregating at transport layer seems most attractive.

\subsection{Decoupling helps overcoming the weakness of asymmetric links}
  %\vspace{-.5em}
The coupled design of downlink and uplink brings the weakness of asymmetric. Asymmetries of links impact the performance of transfer protocol~\cite{asym-2001}. A typical problem of TCP cause by link asymmetries is data packets transfer rate is limited by the ACK packet transfer rate in opposite direction. Many variants of TCP is proposed to alleviate the drawback cause by ACK. There are selective ACK, block ACK, ACK compression, etc.
Another problem is data pendulum, a phenomenon happening when two antiparallel TCP connections share an asymmetric bottleneck link. In these conditions, data and ACK segments alternatively fill only one of the bottleneck link buffer at a time, but almost never both of them. Various solutions have been proposed to handle it~\cite{asym-2016}.
Fortunately, most terminals are now equipped with multiple network interfaces. With the concept of TCP decoupling, data packets and ACK packets can be transferred on two different interfaces. This will change the link asymmetries. If multiple paths can be combined, the weakness of asymmetries may be alleviated to some extent.

\vspace{.6em}
\section{How to decouple the two directions of TCP}
\vspace{.6em}
  %\vspace{-.5em}
In traditional networks, interfaces of network nodes are configured as duplex. A bidirectional link can be established between interfaces of source node and destination nodes. Consequently, a TCP connection can be established over concatenated bidirectional links. With coupled design of links on two directions, the two directions of TCP path are coupled also.
In fact, a unidirectional interface is easier to design with lower cost in some circumstances. Today, most smart terminals are equipped with multiple interfaces, which are bidirectional or unidirectional. It is possible to create unidirectional or bidirectional links over such interfaces. A path consists of few concatenated links may be unidirectional or bidirectional. In this section, the intrinsic relations between interface, link, path and IP address are discussed in details. Then the principle of TCP decoupling is described through variety scenarios with unidirectional paths.

\subsection{Duplex interface and simplex interface}
  %\vspace{-.5em}
A terminal connected to network should be equipped at least one network interface. A network interface card (NIC) is a circuit module that is integrated or packaged in the terminal. Traditionally, an interface is duplex, which can send packets and received packets simultaneous or unsimultaneous. Even more, most terminals have only on interface in the past decades. And then, transport layer protocols assume that the link established on a single duplex interface.
It's worth noting that not all interfaces are duplex. For example, the communication interface of a GPS receiver is simplex, which can only receiving. Considering high power electromagnetic radiation is hazardous to human, some communication technologies is not suitable to handled mobile terminals. Although many emerging technologies such as visible light communications, millimeter-wave communications, and ultraviolet communications introduce high rate, emitting at such frequencies on a handled terminal brings risk of health. A more safe and feasible method to utilize these communication technologies on handled terminals is to integrate a simplex interface which can only receiving.

\subsection{Unidirectional or bidirectional links}
\begin{figure*}[!t]
 \centering
 \includegraphics[scale = 0.4]{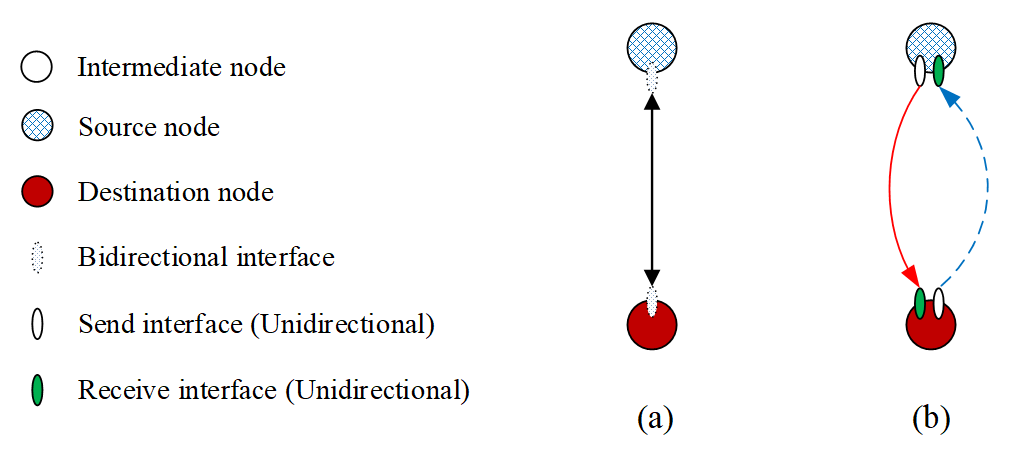}
\caption{Interfaces connections on a simplest path with single link}
 \label{fig6}
\end{figure*}

  %\vspace{-.5em}
A link between two interfaces offers the connection between two nodes. A link is bidirectional if the two connected interfaces are duplex. However, a unidirectional link consists of two simplex interfaces, a sending only interface and a receiving only interface. A simplest path consist of only single link between two network nodes. In scenario in Fig. \ref{fig6}(a), the path is established on a bidirectional link over two bidirectional interfaces of source node and destination node. In contrast, the path in Fig. \ref{fig6}(b), is established on two unidirectional link because all the interfaces are unidirectional.

\subsection{Path over unidirectional or bidirectional links}
\begin{figure*}[!t]
 \centering
 \includegraphics[scale = 0.4]{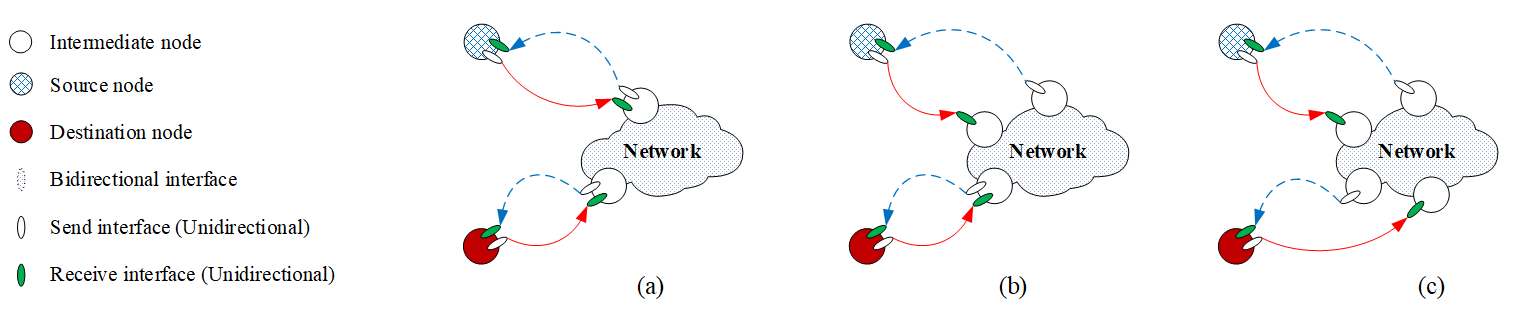}
\caption{Path between two nodes consists of variety links}
 \label{fig7}
\end{figure*}

  %\vspace{-.5em}
Actually, path between the source node and the destination node can be more variety. Let's begin with three possible scenarios. The first, in Fig. \ref{fig7}(a), the source node connected to core network through a intermedia node (default gateway of its subnet), meanwhile, the destination nodes is similar. All interfaces on each node are duplex. The second, if the source node is configured with two unidirectional interfaces, two possible unidirectional links are shown in Fig. \ref{fig7}(b). There are slight differences at the destination node. It lies in that the next hop of the destination node is configured with two unidirectional interfaces too. The third, all intermediate nodes in the network are configured with unidirectional interface only in Fig. \ref{fig7}(c).

\subsection{IP addresses configuration of duplex/simplex interfaces}
\begin{figure*}[!t]
 \centering
 \includegraphics[scale = 0.4]{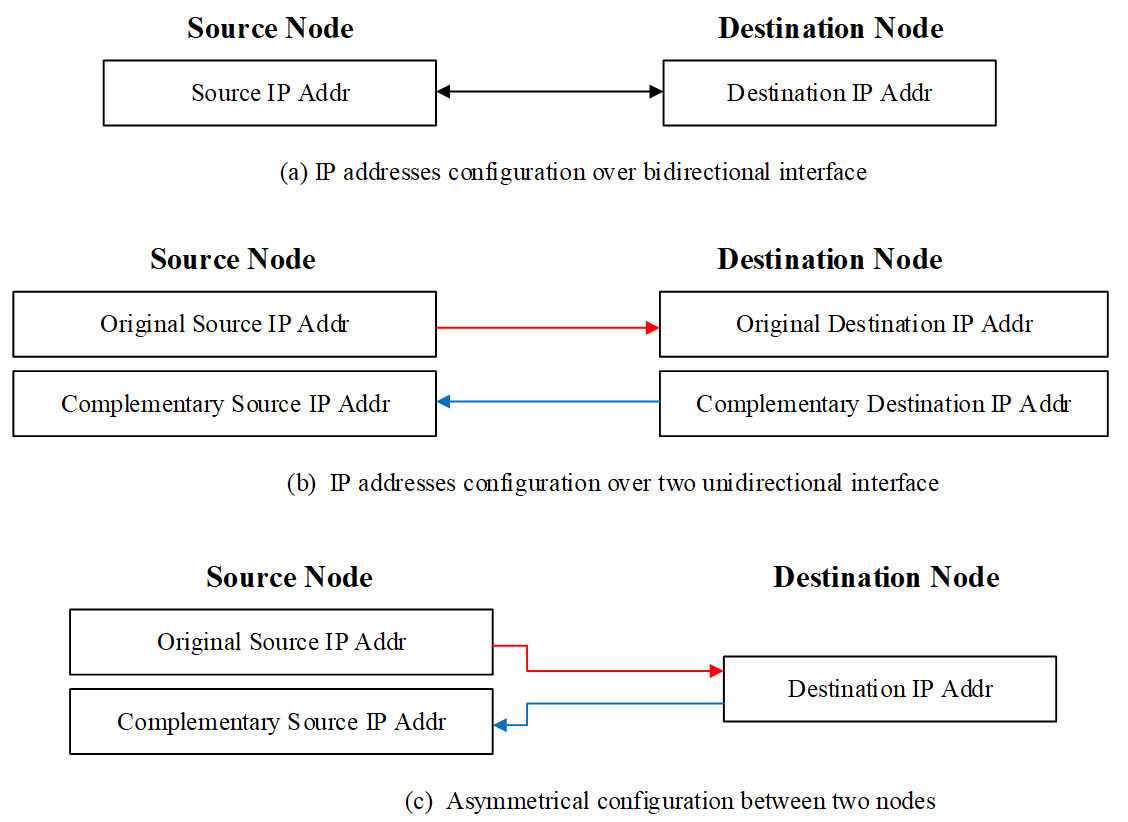}
\caption{Use cases of interfaces and their IP addresses}
 \label{fig8}
\end{figure*}

  %\vspace{-.5em}
Each interface is bound with a MAC address at link layer. A MAC address is mapped to a certain IP address by ARP protocol. When a node (source) wants to send data to another node (destination), we need to decide source IP address and destination IP address. The source IP address indicates which interface should be used to send or receive packets. Traditional terminal holds only one active network interface which is bidirectional. So there are only one source IP address and one destination IP address, as shown in Fig. \ref{fig8} (a).
Refer to a terminal with two simplex (unidirectional) interfaces, two addresses are needed. For convenience, we denote the IP address of sending interface at source node as 'original source IP address', the IP address of receiving interface at source node as 'complementary source IP address'. Meanwhile, we denote the IP address of receiving interface at destination node as 'original destination IP address', the IP address of sending interface at destination node as 'complementary destination IP address'. These four IP addresses at source node and destination node are shown in Fig. \ref{fig8} (b). Two unidirectional paths are established, they are from original source IP address to original destination IP address, from complementary destination IP address to complementary source IP address.
Use case shown in Fig. \ref{fig8} (c) can be seen as the degradation of Fig. \ref{fig8} (b). In this scenario, one of the nodes is configured with two unidirectional interfaces, and, the other is configured with bidirectional interface. There are two unidirectional paths, from original source IP address to destination IP address, from destination IP address to complementary source IP address.

\vspace{.6em}
\section{Implementation of decoupled TCP}
\vspace{.6em}
  %\vspace{-.5em}
In standard TCP protocol, each connection is identified by a four-tuple including source address, source port, destination address, and destination port. The premise behind this control mechanism is that TCP connection is created on a bidirectional logical path. Now that decoupled TCP needs cooperation of two unidirectional paths, a six-tuple (original source address, complementary source address, source port, original destination address, complementary destination address, destination port) is necessary to specify a decoupled TCP connection. Packet transmission of decoupled TCP connection will be indicated by the six-tuple.
To satisfy the second goal, the structure of socket remains unchanged, we look for other solutions to pass the complementary address into protocol stack. Kernel configuration file is an obvious and effective way to realize the above function. In the design of our decoupled TCP, users can set the complementary address with configuration file, so that the complete six-tuple is created with the original standard sockets pair and complementary addresses. In this way, decoupled TCP guarantees the compatibility with application layer.

\subsection{Connection Initiation and Close}
\begin{figure*}[!t]
 \centering
 \includegraphics[scale = 0.5]{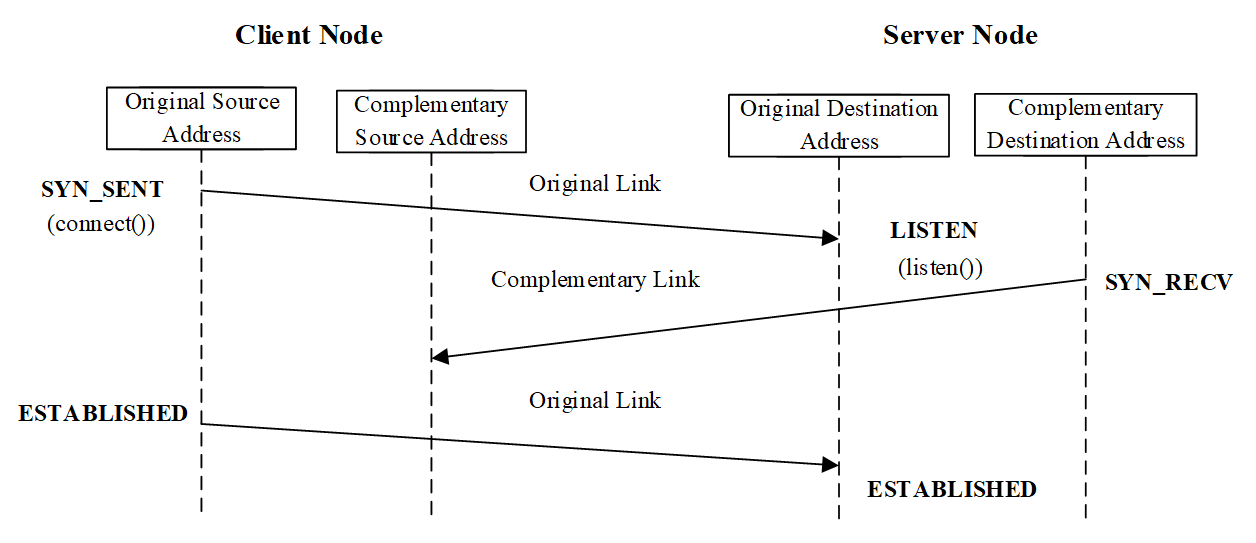}
\caption{Three-way handshake directed by six-tuple in connection initiation phase}
 \label{fig9}
\end{figure*}

\begin{figure*}[!t]
 \centering
 \includegraphics[scale = 0.5]{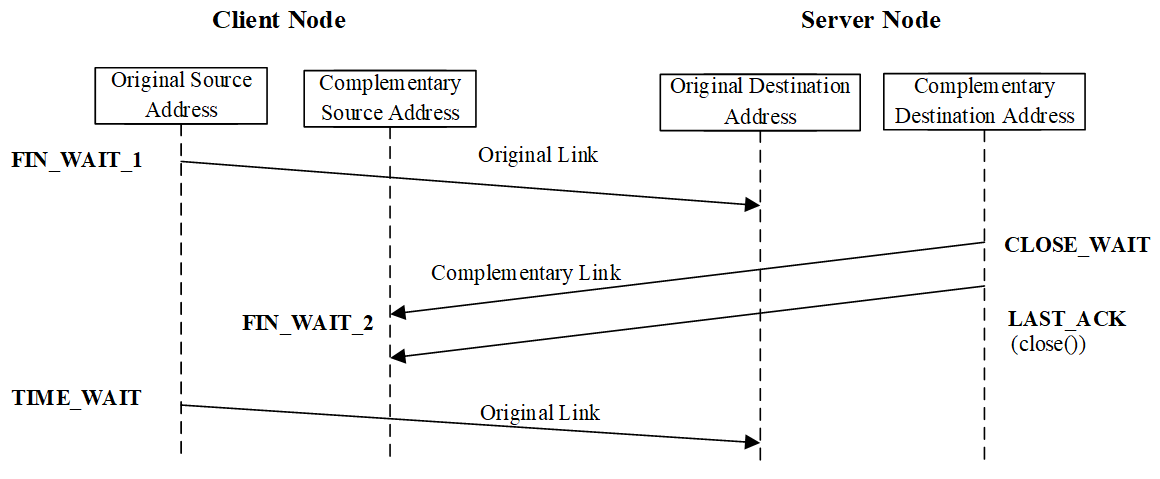}
\caption{Four-way handshake of decoupled TCP connection close}
 \label{fig10}
\end{figure*}

  %\vspace{-.5em}
The process of initiating and closing of decoupled TCP connection is quite similar to the normal TCP's. As a maturity and effective method, the handshake mechanism ~\cite{tcp-78} is remained, but there are two main changes between decoupled TCP and standard TCP in the initiating and closing stage:
Because local complementary address is unknown to the remote host before connection establishment, and the complete address information is necessary for each host to execute decoupled transmission, we add a TCP option to inform remote host about complementary address in connection initiation process.
The trait of decoupling shows up in both connection initiation stage and connection close stage. It particularly reflects on transmission interface handover according to the interface's transmission direction set before. For example, Fig. \ref{fig9} shows the three-way handshake initiation process directed by the six-tuple. Fig. \ref{fig10} shows the four-way handshake termination process of decoupled TCP.

\subsection{Data Transfer}
\begin{figure*}[!t]
 \centering
 \includegraphics[scale = 0.5]{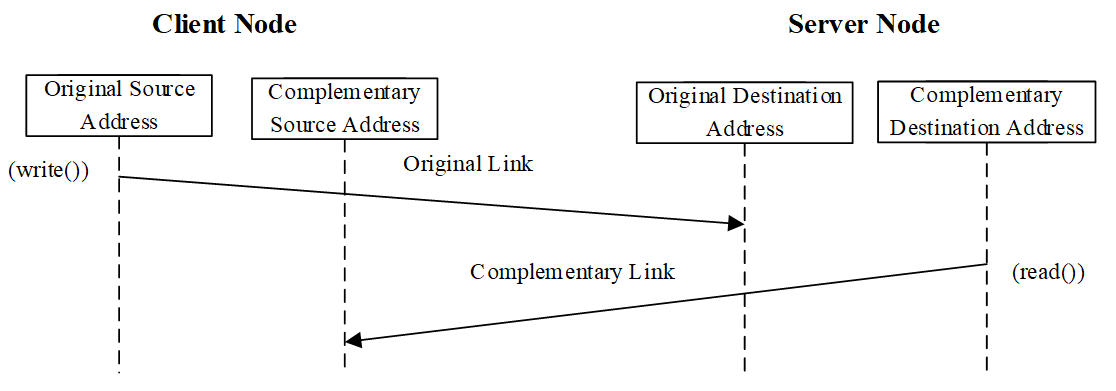}
\caption{Packets transfer phase of decoupled TCP}
 \label{fig11}
\end{figure*}

  %\vspace{-.5em}
In order to ensure the decoupling in data transfer process, when the next packet (data or ACK) is ready to be sent, the current interface's direction must be checked. If it's direction is from remote host to local host, the sending work will be accomplished on its complementary link. The procedure is shown in Fig. \ref{fig11}.
In the process of data reception, there is yet another significant problem to be addressed. If kernel uses address parsed from IP header directly to demultiplex arriving packets, the packets may be bound to wrong unidirectional connection. For example, server sends data packet to client with downlink, and the client will echo back ACK with uplink. In this situation, addresses parsed from ACK packet is uplink's address pair, but for server, connection is bound to downlink's addresses, and this mismatching leads to misoperation to incoming packet. We add a new TCP option to handle this problem.

\subsection{States of decoupled TCP}
\begin{figure*}[!t]
 \centering
 \includegraphics[scale = 0.5]{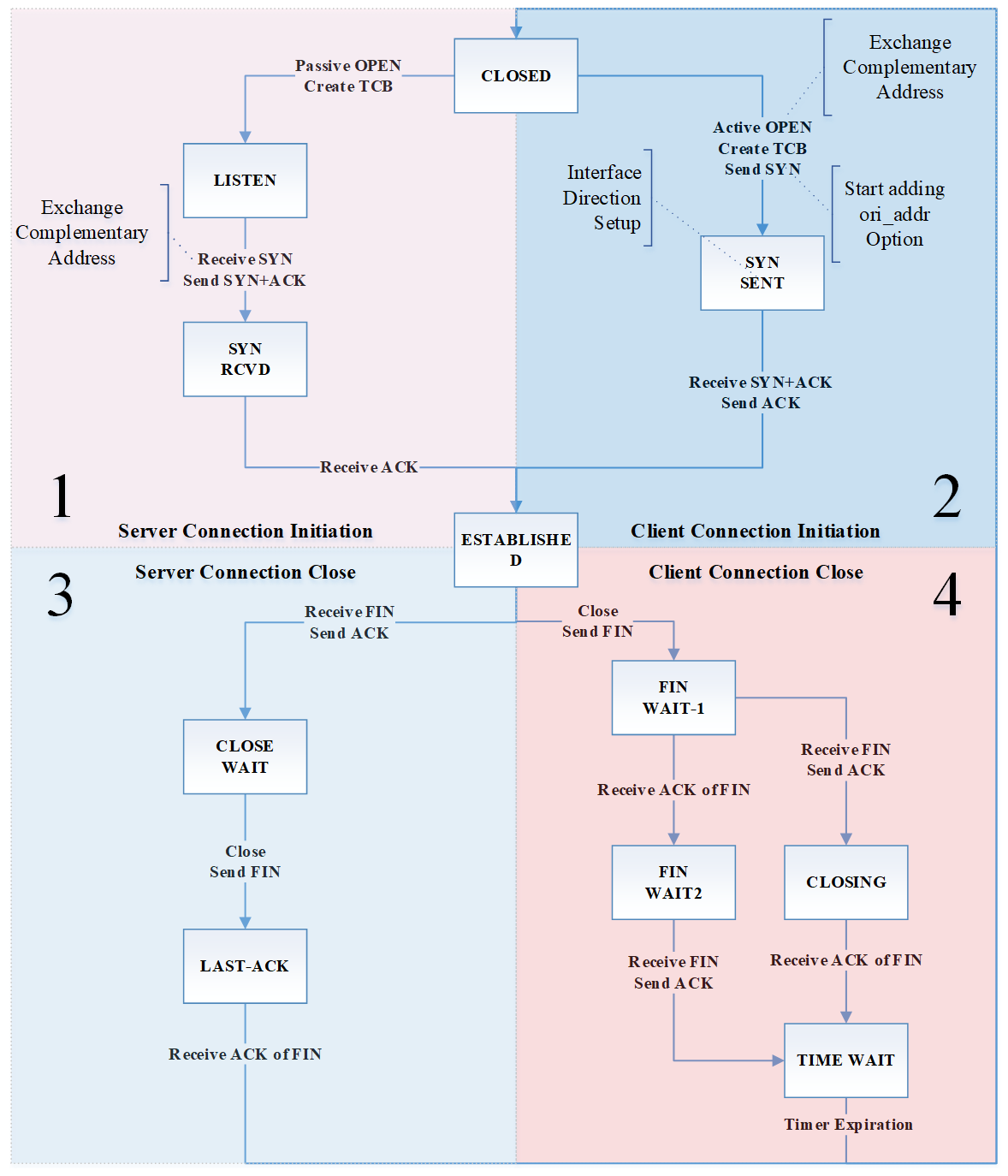}
\caption{State diagram of decoupled TCP}
 \label{fig12}
\end{figure*}

  %\vspace{-.5em}
The state diagram of decoupled TCP is shown in Fig. \ref{fig12}. The left half is server's overall process, and the part of client is shown in the right half of the figure. We annotate decoupled TCP's major changes on the figure: In the stage of connection initiation, apart from conventional decoupling to transmission, several settings and options help connection to establish correctly; after laying the groundwork, packets will be transferred on appropriate link according to their direction.

\subsection{Availability Testing}
  %\vspace{-.5em}
We have implemented a decoupled TCP in Linux kernel. An experiment over VLC hybrid network has been carried out. The decoupled TCP is tested on combinations of VLC downlink and Ethernet uplink. Performance under different packet loss rates have been analyzed also.
The results show that the decoupled TCP achieves highly stable and efficient transmission in VLC hybrid system without packet loss. In this scenario, the utilization ratio of the bandwidth reaches 95\%. When there are packet losses, the throughput remains stable with packet loss ratios of 0.5\%, 1.0\% or 2.0\% and the utilization ratio of the bandwidth can still reach over 92\%. Only when the packet loss ratio reaches over 5\%, the performance degrades obviously.

\vspace{.6em}
\section{Open Issues}
\vspace{.6em}
Idea of aggregating is boosting variety skills to utilize multiple communication resources simultaneously. Although many researchers pay their attention to aggregate at link layer~\cite{dual-2016}~\cite{dual-2017}, decoupling of TCP provides a novel method to aggregate bidirectional or unidirectional interfaces at transport layer. Our experiment has proved the availability of decoupled TCP. How to improve the performance of it under heterogeneous networks needs further study.

Today, most smart terminals are equipped with multiple interfaces. To address multi-homing problem, IETF proposed MPTCP~\cite{mptcp-2013} which allows a single connection to transmit packets on multiple paths simultaneously. Some research has shown that MPTCP benefits to aggregate WiFi and 4G ~\cite{handover-2012}. How to combine the idea of decoupling and multipath at transport layer needs further discussion.

% BIBLIOGRAPHY
%\addcontentsline{toc}{chapter}{Bibliography}
\vspace{.6em}
\bibliographystyle{ieeetr}
\bibliography{DecoupledTCP}

\end{document}

%% file: Macros.tex
\def\bf{{\pmb f}}

\newcommand{\upperRomannumeral}[1]{\uppercase\expandafter{\romannumeral#1}}